\documentclass{ws-ijmpd}
\usepackage[super,compress]{cite}
\usepackage{graphicx}
\usepackage{bm}
\usepackage[mathlines]{lineno}
\usepackage{tensor}
\usepackage{amsmath,amsfonts,mathtools}
\usepackage{textcase}

\begin{document}
\markboth{Kian Ming, Triyanta, J. S. Kosasih}
{Gravitoelectromagnetism in Teleparallel Equivalent of General Relativity: A New Alternative}
	
%
\catchline{}{}{}{}{}
%
	
\title{Gravitoelectromagnetism in Teleparallel Equivalent of General Relativity: A New Alternative}

\author{KIAN MING}
\address{Department of Physics, Parahyangan Catholic University, Jalan Ciumbuleuit 94, Bandung 40112, Indonesia\\ Division of Theoretical High Energy Physics and Instrumentation, Faculty of Mathematics and Natural sciences, Institut Teknologi Bandung, Jalan Ganesha 10, Bandung 40132 Indonesia\\ Indonesian Center for Theoretical and Mathematical Physics\\ kian.ming@unpar.ac.id}
	
\author{TRIYANTA}
\address{Division of Theoretical High Energy Physics and Instrumentation, Faculty of Mathematics and Natural sciences, Institut Teknologi Bandung, Jalan Ganesha 10, Bandung 40132 Indonesia\\ Indonesian Center for Theoretical and Mathematical Physics\\triyanta@fi.itb.ac.id }
	
\author{J. S. KOSASIH}
\address{Division of Theoretical High Energy Physics and Instrumentation, Faculty of Mathematics and Natural sciences, Institut Teknologi Bandung, Jalan Ganesha 10, Bandung 40132 Indonesia\\ Indonesian Center for Theoretical and Mathematical Physics\\jusak@fi.itb.ac.id}

\maketitle
	
\begin{history}
\received{Day Month Year}
\revised{Day Month Year}
\end{history}
	
\begin{abstract}
Spaniol and Andrade introduced grvitoelectromagnetism in TEGR by considering superpotentials, times the determinant of tetrads, as the gravitoelectromagnetic fields. However, since this defined gravitoelectromagnetic field strength does not give rise to a complete set of Maxwell-like equations, we propose an alternative definition of the gravitoelectromagnetic field strength: instead of superpotentials, torsions are taken as the gravitoelectromagnetic field strengths. Based on this new proposal we are able to derive a complete set of Maxwell-like equations. We then apply them to obtain explicit expressions of the gravitoelectromagnetic fields both in Schwarzchilds spacetime and for gravitational waves.
		
\end{abstract}
	
\ccode{PACS numbers:04.20.Cv, 04.50.-h}
	
	
\section{Introduction}
	
Stationary electric charges generate electric fields. When the source charge moves with a constant velocity, or equivalently when we look at an inertial frame that moves with a constant velocity with respect to the charge, we in addition also detect magnetic fields. The similar expression of the Newton's law of gravitation as the Coulomb's law of electricity leads one naturally to define an electric and magnetic field analog of gravitation [\citen{Mashhoon2001,Mashhoon2008,Ulrych2006, Heavi1893}], called the gravitoelectric and gravitomagnetic fields, or the gravitoelectromagnetic field. The gravitational Lorentz force can then be defined where its velocity dependent part (also acceleration dependent part for a more general force) corresponds to a gravitomagnetic field. This magnetic part of the gravitational force leads the solar planetary orbits to precess [\citen{Gine2006}]. In fact, the gravitomagnetic version of the Einstein’s general relativity, for the case of weak field and low velocity of the source, predicts more accurately the precession of the solar planetary orbits [\citen{Gine2006,Spaniol2010}]. Therefore, gravitoelectromagnetism (GEM) is one important formalism of gravitation.

Theory of GEM was introduced due to the similar expression between the Coulomb force and the Newton gravitational force. Accordingly, as the electromagnetism is governed by the Maxwell’s equations, the Newtonian GEM is governed by the Maxwell-like gravitational equations. Newtonian gravitational force is equivalent to the Coulomb force through the interchange of mass $\leftrightarrow$ charge and $-\textit{G }\leftrightarrow 1/{4 \pi \epsilon_{0}}$. The mass of an object cannot be negative; the charge on the other hand can have either a positive or a negative value. Accordingly, the Coulomb force can be attractive or repulsive while the gravitational force should be attractive. The equivalence of both forces leads one to introduce magnetic and electric fields analog of gravitation (the gravitomagnetic field $ \vec{B}_{g}$ and gravitoelectric field $\vec{E}_g$) through the Maxwell-like equations
\begin{equation}
\begin{split}
& \nabla \cdot \vec{E}_{g} =-4\pi G\rho _{g},  \\ 
& \nabla \times \vec{B}_{g} -\frac{1}{c^{2}} \frac{\partial \vec{E}_{g} }{\partial t} =\frac{-4\pi G}{c^{2} } \vec{J}_{g}, \\
& \nabla \cdot \vec{B}_{g} =0,\\
& \nabla \times \vec{E}_{g} +\frac{\partial \vec{B}_{g} }{\partial t} =0. \label{eq:M1}
\end{split}
\end{equation}
They differ only by some constant factors. In fact, interchanging $4 \pi G \rightarrow 1/{\epsilon _0}$ makes the above equations are identical to Maxwell's equations.
	
Any test mass \textit{m} moving with a velocity $\vec{v}$ in the presence of gravitoelectromagnetic fields $(\vec{E}_{g} ,\vec{B}_{g} )$ experiences a gravitational Lorentz force $\vec{F}_{g} =m\gamma (v)[\vec{E}_{g} +\vec{v}\times \vec{B}_{g} ]$. Recalling that the first term of the gravitational Lorentz force leads to stationary elliptical orbit of planets, the appearance of the second term, the velocity dependence term, will change such a stationary orbit. Before the birth of general relativity, we have the fact that the Mercury's perihelion moves [\citen{Mashhoon2000}]. This fact lead to the proposal of the need of extra terms in the gravitational force. And in fact there were some proposals for extra terms including [\citen{Mashhoon2008,Gine2009,Holzmuller1870,Aldrovandi2013,Tisserand1872,Tisserand1890,Gerber1917,Weber2004}]. The proposals, however, could not explain the Mercury's orbit precession as better as the Einstein's proposal of general relativity.
	
Einstein's field equation in general relativity is non-linear. So to obtain gravitoelectromagnetic fields fulfilling the Maxwell-like equations for this theory we should consider the weak field case [\citen{Mashhoon2008,Medina2006}], i.e. we should take a linear approximation. In this case, the metric tensor ${g_{\mu \nu }}$ is close to the Minkowski metric tensor ${\eta _{\mu \nu }}$ [\citen{Ryder1996}]
\begin{equation}
{g_{\mu \nu }} = {\eta _{\mu \nu }} + {h_{\mu \nu }},\quad {g^{\mu \nu }} = {\eta ^{\mu \nu }} - {h^{\mu \nu }},\,|{h_{\mu \nu }}| << 1. \label{eq:AP}
\end{equation}
	
They are two different definitions of gravitoelectric and gravitomagnetic fields for the Einstein's gravity. First, the fields are defined through a similar relationship between electromagnetic fields and potentials in electromagnetism where $\overline{h}_{\mu \nu}={h_{\mu \nu }}-{\tfrac{1}{2}} h {\eta _{\mu \nu }}$  are taken as the gravito-potentials. Second, the fields are defined as the components of the Weyl tensors $C_{\mu\nu\alpha\beta}$ [\citen{Ramos2010,Campbell1971,Campbell1973,Campbell1976}]
\begin{equation}
E_{ij} =-c^{2} C_{0i0j} , B_{ij} ={\tfrac{1}{2}} c\varepsilon _{ikl} \tensor{C}{^{kl}_{0j}} . \label{eq:EGE1}
\end{equation}
Using expressions of the Weyl tensor in terms of Riemannian tensors and their symmetry properties, one can obtained the Maxwell-like equations
\begin{equation*}
\begin{split}
& \nabla \cdot \vec{E}_{j} \approx -{\tfrac{4\pi G}{c^{2} }} \left(T_{00,j} + \tensor{T}{_{jl,}^{l}} \right)\, \, +{\tfrac{4\pi G}{3c^{2} }} T_{,j} ,\\
& \nabla \cdot \vec{B}_{j} \approx \varepsilon _{kjn} \eta ^{ns} {\tfrac{4\pi G}{c^{3} }} \tensor{T}{_{s0,}^{k}} ,\\ 
& [(\nabla \times \vec{E}_{j} )+c\partial _{0} \vec{B}_{j} ]_{k} \approx {\tfrac{8\pi G}{3c^{2} }} \varepsilon _{klj} \partial ^{l} T-{\tfrac{4\pi G}{c^{2} }} \varepsilon _{klm} \partial ^{m} T^{l} _{j} ,\\
& [(\nabla \times \vec{B}_{j} )-{\tfrac{1}{c}} \partial _{0} \vec{E}_{j} ]_{k} \approx -{\tfrac{4\pi G}{c^{3} }} \left(T_{0k;j} - T_{jk;0} \right) -{\tfrac{4\pi G}{3c^{3} }} \eta _{jk} T_{;0},
\end{split}
\end{equation*}
where $T_{\alpha \beta}$ are energy-momentum tensors. We identify from the above equations that the gravito charge density and gravito current density are
\begin{equation*}
\begin{split}
& \rho _{j} \approx {\tfrac{1}{c}} (T_{00,j} + \tensor{T}{_{jl,}^{l}}) -{\tfrac{1}{3c}} T_{,j} ,\\
& (\vec{J}_{j} )_{k} \approx {\tfrac{1}{c}} (T_{0k;j} -T_{jk;0}) +{\tfrac{1}{3c}} \eta _{jk} T_{;0} .
\end{split}
\end{equation*}
Here the source terms are quite complicated and depend on the fields themselves. As electromagnetism (EM) is governed by the Maxwell's equations, the GEM is equivalently governed by equations that have similar forms as the Maxwell's equations, the Maxwell-like equations of gravity. 
	
Unlike EM, theory of gravity is not a unique theory. We have Newtonian gravity as well as the Einstein gravity (Einstein general relativity) with some variants of it. In addition, there is also teleparallel gravity (TG) or teleparallel equivalent of general relativity (TEGR) [\citen{Andrade1997,Aldrovandi2013}], a gauge theory of gravity with translational group as the gauge symmetry. A definition of gravitoelectromagnetic fields in TEGR has been proposed by Spaniol and Andrade [\citen{Spaniol2010}]. Unfortunately their definition does not lead to the field strength fulfilling the Bianchi identity in a usual form. In this paper we propose a different definition of gravitoelectromagnetic fields that gives rise to a complete set of Maxwell-like equations. The organization of the paper is the following. The second part is given to review the concept of TEGR especially on GEM formulation in TEGR scheme by Spaniol and Andrade. Next in the third part we propose a new alternative definition of GEM in TEGR scheme. Finally, before conclusions, we apply our GEM for Schwarzschild metric and for gravitational waves.

\section{Spaniol-Andrade's GEM in TEGR}

TG or TEGR is an alternative way of describing gravity. It is a gauge theory for translational group [\citen{Spaniol2010,Aldrovandi2013}]. As the translational group is abelian, the corresponding field strength is similar as that of the Maxwell field
\begin{equation}
\begin{split}
\tensor{F}{^{a}_{\mu \nu }} & \equiv \tensor{T}{^{a} _{\mu \nu }} =\partial _{\mu } \tensor{b}{^{a} _{\nu }} -\partial _{\nu } \tensor{b}{^{a} _{\mu }} \\
& =\partial _{\mu } \tensor{h}{^{a} _{\nu }} -\partial _{\nu } \tensor{h}{^{a} _{\mu }}\\ 
& =\tensor{h}{^{a} _{\rho }} \tensor{T}{^{\rho } _{\mu \nu }} . \label{eq:TT}
\end{split}
\end{equation}
	
\noindent In the above, Greek and Latin indices correspond to curved spacetime and its tangent space (Minkowski spacetime) respectively. We will use middle Latin alphabets $i,j,k$ for space components of tensors in curved spacetime. The translational gauge potential $\tensor{b}{^{a}_{\mu}}$ is defined to relate the tetrad field $\tensor{h}{^{a}_{\mu}}$ through
\begin{equation}
\tensor{h}{^{a}_{\mu}}=\partial _{\mu } x^{a} +\tensor{b}{^{a}_{\mu}}. \label{eq:H}
\end{equation} 
\noindent It corresponds to the metric tensor through
\begin{align}
g_{\mu \nu } =\eta _{ab} \tensor{h}{^{a}_{\mu }} \tensor{h}{^{b}_{ \nu }} \label{eq:S1}\\
\eta _{ab} =g_{\mu \nu } \tensor{h}{_{a}^{\mu}} \tensor{h}{_{b}^{\nu}}. \label{eq:S2}
\end{align}
\noindent In TEGR, gravitational fields are represented by the tetrad fields.
	
TEGR is defined in a curvature-less Weitzenb\"{o}ck space, a spacetime characterized by the Weitzenb\"{o}ck connection $\tensor{\Gamma}{^{\rho}_{\mu \nu}}=\tensor{h}{_{a}^{\rho}}\partial_{\mu}\tensor{h}{^{a}_{\nu}}$. The gravitational field strenght is fully represented by the torsion $\tensor{T}{^{\rho}_{\mu\nu}}=\tensor{\Gamma}{^{\rho}_{\nu\mu}}-\tensor{\Gamma}{^{\rho}_{\mu\nu}}$. The Lagrangian of the gravitational field is [\citen{Aldrovandi2013}]
	
\begin{equation}
\mathfrak{L}=\frac{h}{16\pi G} S^{\rho \mu \nu } T_{\rho \mu \nu } \label{eq:L} 
\end{equation} 
which is equivalent to the Lagrangian of the gravitational filed in GR up to a total derivative [\citen{Aldrovandi2013}]. The corresponding field equation is
\begin{equation}
\partial_{\rho}(h \tensor{S}{_{a}^{\rho \sigma}}) = 4\pi G (h \tensor{J}{_{a}^{\sigma}}) .\label{eq:A}
\end{equation}
	
\noindent In the above  $h=det(\tensor{h}{^{a}_{\mu}})$, and
	
\begin{equation}
S^{\rho \mu \nu } ={\tfrac{1}{2}} (K^{\mu \nu \rho } -g^{\rho \nu } \tensor{T}{^{\alpha \mu } _{\alpha}} +g^{\rho \mu} \tensor{T}{^{\alpha \nu}_{\alpha}} ) , \label{eq:D}
\end{equation}
\begin{equation}
\tensor{K}{^{\rho}_{\mu \nu}} ={\tfrac{1}{2}}(\tensor{T}{_{\nu}^{\rho} _{\mu}} + \tensor{T}{_{\mu}^{\rho}_{\nu}} - \tensor{T}{^{\rho}_{\mu \nu}} ),  \label{eq:E} 
\end{equation}
\begin{equation}
\tensor{J}{_{a}^{\rho}} =\frac{\partial {\mathfrak{L}}}{\partial h^{a} _{\rho } } =\frac{\tensor{h}{_{a} ^{\lambda}}}{4\pi G} (\tensor{T}{^{c}_{\mu \lambda}}\tensor{S}{_{c} ^{\mu \rho }} -{\tfrac{1}{4}} \tensor{\delta}{ _{\lambda} ^{\rho}} \tensor{T}{^{c}_{\mu \nu}} \tensor{S}{_{c} ^{\mu \nu}} ).  \label{eq:F}
\end{equation}
	
Equation (\ref{eq:A}) reminds us to the Maxwell's equations 
\begin{equation}
\begin{split}
& \partial _{\mu } F^{\mu \nu } =J^{\nu } \label{eq:MW6}
\end{split}
\end{equation}
with $h\tensor{S}{_{a}^{\rho \sigma}}$ act as the (gravitational) field strengths. In addition, $h\tensor{S}{_{a}^{\rho \sigma}}$ are anti-symmetric, $h\tensor{S}{_{a}^{\rho \sigma}} = -\tensor{S}{_{a}^{\sigma \rho}}$, just like the electromagnetic field strengths. This property is due to the anti-symmetry of the torsion. Unlike to the original Maxwell's equations, $\tensor{J}{_{a}^{\rho}}$ in the above Maxwell's like equations do not represent external sources as they correspond only to the gravitational field. This tensor also represents the Noether energy-momentum density [\citen{Aldrovandi2013}]. Considering the similarity between (\ref{eq:A}) and (\ref{eq:MW6}) Spaniol and Andrade introduced  gravitoelectric and gravitomagnetic fields as follows [\citen{Spaniol2010}] 
\begin{equation}
h\tensor{S}{_{a}^{0i}} = \tensor{E}{_{a} ^{i}} , \quad
h\tensor{S}{_{a} ^{ij}} = c\varepsilon ^{ijk} B_{ak}. \label{eq:A2} 
\end{equation} 
\noindent The later is equivalent to
\[\tensor{B}{_{a} ^{i}} =-{\tfrac{1}{3c}} \varepsilon ^{ijk} hS_{ajk} .\] 
	
The equation (\ref{eq:A}) then gives the Gauss' law and the Ampere's law for gravity
\begin{equation}
\begin{split}
& \nabla \cdot \vec{E}_{a} =-4\pi G(h \tensor{J}{_{a} ^{0}} ), \\
& \frac{1}{c^{2} } \frac{\partial }{\partial t} \vec{E}_{a} +\nabla \times \vec{B}_{a} =\frac{4\pi G}{c} (h\vec{J}_{a} ).
\end{split} 
\end{equation}
	
In addition, the torsion can be expressed in terms of field strengths and tetrads according to [\citen{Spaniol2010}]
\begin{equation} 
\begin{split}
\tensor{T}{^{c}_{\gamma \delta}} = &  2\tensor{h}{^{b} _{\gamma}} g_{\alpha \delta } \tensor{h}{^{c}_{\sigma}} \tensor{S}{_{b} ^{\sigma \alpha}} -2\tensor{h}{^{b} _{\delta}} g_{\alpha \gamma } \tensor{h}{^{c} _{\sigma}} \tensor{S}{_{b} ^{\sigma \alpha}}\\
& - \tensor{h}{^{c} _{\delta}} g_{\alpha \gamma } \tensor{h}{^{b} _{\sigma}} \tensor{S}{_{b} ^{\sigma \alpha}} + \tensor{h}{^{c} _{\gamma}} g_{\alpha \delta } \tensor{h}{^{b} _{\sigma}} \tensor{S}{_{b} ^{\sigma \alpha}} . \label{eq:B}
\end{split}
\end{equation}

\noindent Accordingly we have
\begin{equation*}
\begin{split}
4\pi G & \tensor{J}{_{a} ^{\rho }}  =  \tensor{h}{_{a} ^{\lambda }} ( 2\tensor{h}{^{b} _{\mu }} g_{\alpha \lambda } \tensor{h}{^{c} _{\sigma }} \tensor{S}{_{b} ^{\sigma \alpha}} \tensor{S}{_{c} ^{\mu \rho}}
- 2\tensor{h}{^{b} _{\lambda}} g_{\alpha \mu } \tensor{h}{^{c} _{\sigma}} \tensor{S}{_{b} ^{\sigma \alpha}} \tensor{S}{_{c} ^{\mu \rho }} 
-\tensor{h}{^{c} _{\lambda}} g_{\alpha \mu } \tensor{h}{^{b} _{\sigma }} \tensor{S}{_{b} ^{\sigma \alpha}} \tensor{S}{_{c} ^{\mu \rho}} \\
&+ \tensor{h}{^{c} _{\mu}} g_{\alpha \lambda } \tensor{h}{^{b} _{\sigma}} \tensor{S}{_{b} ^{\sigma \alpha}} \tensor{S}{_{c} ^{\mu \rho }}
-{\tfrac{1}{4}} \tensor{\delta}{ _{\lambda } ^{\rho}} (2\tensor{h}{^{b} _{\mu }} g_{\alpha \nu } \tensor{h}{^{c} _{\sigma}} \tensor{S}{_{b} ^{\sigma \alpha}} \tensor{S}{_{c} ^{\mu \nu}} 
- 2\tensor{h}{^{b} _{\nu}} g_{\alpha \mu } \tensor{h}{^{c} _{\sigma}} \tensor{S}{_{b} ^{\sigma \alpha}} \tensor{S}{_{c} ^{\mu \nu }} \\
& - \tensor{h}{^{c} _{\nu}} g_{\alpha \mu } \tensor{h}{^{b} _{\sigma }} \tensor{S}{_{b} ^{\sigma \alpha}} \tensor{S}{_{c} ^{\mu \nu}} 
+ \tensor{h}{^{c} _{\mu}} g_{\alpha \nu } \tensor{h}{^{b} _{\sigma}} \tensor{S}{_{b} ^{\sigma \alpha}} \tensor{S}{_{c} ^{\mu \nu}} ) ) .
\end{split}
\end{equation*}

It turns out that the sources $\tensor{J}{_{a}^{\rho}}$ fully depend on gravitational field and do not correspond to external sources. Note also that $\tensor{J}{_{a}^{\rho}}$ are non-linear in field strengths. 
Expressing $\tensor{J}{_{a}^{\rho}}$ explicitly in gravitoelectric $E$ and gravitomagnetic fields $B$, after some algebra, gives the complete Gauss-like and the Ampere-like equations:
\begin{equation*}
\begin{split}
\nabla {\cdot} {\vec{E}_{a}} & = {\tfrac{1}{h}} \tensor{h}{ _{a} ^{\lambda}} (2 \tensor{h}{^{b} _{k}} g_{j\lambda } \tensor{h}{^{c} _{0}} + \tensor{h}{^{c} _{k}} g_{j\lambda } \tensor{h}{^{b} _{0}}  
-2 \tensor{h}{^{b} _{\lambda}} g_{jk} \tensor{h}{^{c} _{0}} \\
& - \tensor{h}{^{c} _{\lambda}} g_{jk} \tensor{h}{^{b} _{0}}  + 2 \tensor{h}{^{b} _{k}} g_{0\lambda } \tensor{h}{^{c} _{j}} - \tensor{h}{^{c} _{k}} g_{0\lambda } \tensor{h}{^{b} _{j}} \\
& + 2\tensor{h}{^{b} _{\lambda}} g_{0k} \tensor{h}{^{c} _{j}} + \tensor{h}{^{c} _{\lambda}} g_{0k} \tensor{h}{^{b} _{j}} ) \tensor{E}{_{b} ^{j}} \tensor{E}{_{c} ^{k}}  \\ 
& + {\tfrac{1}{h}} c \tensor{h}{_{a} ^{\lambda}} (2\tensor{h}{^{b} _{l}} g_{k\lambda } \tensor{h}{^{c} _{j}} + \tensor{h}{^{c} _{l}} g_{k\lambda } \tensor{h}{^{b} _{j}}  \\ 
& -2\tensor{h}{^{b} _{\lambda}} g_{kl} \tensor{h}{^{c} _{j}} - \tensor{h}{^{c} _{\lambda}} g_{kl} \tensor{h}{^{b} _{j}} ) \varepsilon ^{jkm} B_{bm} \tensor{E}{_{c} ^{l}} \\
& + {\tfrac{1}{4}} {\tfrac{1}{h}} \tensor{h}{_{a} ^{\lambda}} \tensor{\delta}{ _{\lambda } ^{0}} (2 \tensor{h}{^{b} _{0}} g_{jk} \tensor{h}{^{c} _{0}} -2 \tensor{h}{^{b} _{k}} g_{j0} \tensor{h}{^{c} _{0}} \\ 
& -2 \tensor{h}{^{b} _{0}} g_{0k} \tensor{h}{^{c} _{j}} + 2 \tensor{h}{^{b} _{k}} g_{00} \tensor{h}{^{c} _{j}}) (2\tensor{E}{_{b} ^{j}} \tensor{E}{_{c} ^{k}} + \tensor{E}{_{b} ^{k}} \tensor{E}{_{c} ^{j}} ) \\ 
& + {\tfrac{3}{4}} {\tfrac{1}{h}} c \tensor{h}{_{a} ^{\lambda}} \tensor{\delta}{_{\lambda} ^{0}} (2 \tensor{h}{^{b} _{k}} g_{jl} \tensor{h}{^{c} _{0}} - 2 \tensor{h}{^{b} _{k}} g_{0l} \tensor{h}{^{c} _{j}} \\ 
& - 2 \tensor{h}{^{b} _{0}} g_{kj} \tensor{h}{^{c} _{l}} + 2 \tensor{h}{^{b} _{j}} g_{k0} \tensor{h}{^{c} _{l}} ) \varepsilon ^{klm} \tensor{E}{_{c} ^{j}} B_{bm} \\ 
& + {\tfrac{1}{2}} {\tfrac{1}{h}} c^{2} \tensor{h}{_{a} ^{\lambda}} \tensor{\delta}{_{\lambda} ^{0}} \tensor{h}{^{b} _{l}} g_{km} \tensor{h}{^{c} _{j}} \varepsilon ^{jkp} \varepsilon ^{lmq} (2B_{bp} B_{cq} +B_{bq} B_{cp} ),
\end{split}
\end{equation*}
\noindent
\begin{equation*}
\begin{split}
(\frac{1}{c^{2} } \frac{\partial }{\partial t} \vec{E}_{a} & + \nabla \times \vec{B}_{a})^{i} ={\tfrac{1}{ch}} \tensor{h}{_{a} ^{\lambda}} (2 \tensor{h}{^{b} _{0}} g_{j\lambda } \tensor{h}{^{c} _{0}} + \tensor{h}{^{c} _{0}} g_{j\lambda } \tensor{h}{^{b} _{0}} \\ 
&- 2 \tensor{h}{^{b} _{\lambda}} g_{j0} \tensor{h}{^{c} _{0}} - \tensor{h}{^{c} _{\lambda}} g_{j0} \tensor{h}{^{b} _{0}}  - 2 \tensor{h}{^{b} _{0}} g_{0\lambda } \tensor{h}{^{c} _{j}}\\
& - \tensor{h}{^{c} _{0}} g_{0\lambda } \tensor{h}{^{b} _{j}}  + 2 \tensor{h}{^{b} _{\lambda}} g_{00} \tensor{h}{^{c} _{j}} + \tensor{h}{ ^{c} _{\lambda}} g_{00} \tensor{h}{^{b} _{j}} ) \tensor{E}{_{b} ^{j}} \tensor{E}{_{c} ^{i}} \\
& + {\tfrac{1}{h}} \tensor{h}{_{a} ^{\lambda}} (2 \tensor{h}{^{b} _{k}} g_{j\lambda } \tensor{h}{^{c} _{0}} + \tensor{h}{^{c} _{k}} g_{j\lambda } \tensor{h}{^{b} _{0}}  - 2 \tensor{h}{^{b} _{\lambda}} g_{jk} \tensor{h}{^{c} _{0}} \\
&-\tensor{h}{^{c} _{\lambda}} g_{jk} \tensor{h}{^{b} _{0}}  
+2\tensor{h}{^{b} _{k}} g_{0\lambda } \tensor{h}{^{c} _{j} } - \tensor{h}{^{c} _{k}} g_{0\lambda } \tensor{h}{^{b} _{j}} \\ 
& +2 \tensor{h}{^{b} _{\lambda}} g_{0k} \tensor{h}{^{c} _{j}} + \tensor{h}{^{c} _{\lambda}} g_{0k}\tensor{h}{^{b} _{j}} )\varepsilon ^{kil} \tensor{E}{_{b} ^{j}} B_{cl} \\
& +{\tfrac{1}{h}} \tensor{h}{_{a} ^{\lambda}}  (2 \tensor{h}{^{b} _{0}} g_{k\lambda } \tensor{h}{^{c} _{j}} + \tensor{h}{^{c} _{0}} g_{k\lambda } \tensor{h}{^{b} _{j}} \\ 
& - 2 \tensor{h}{^{b} _{\lambda}} g_{k0} \tensor{h}{^{c} _{j}} - \tensor{h}{^{c} _{\lambda}} g_{k0} \tensor{h}{^{b} _ {j}} )\varepsilon ^{jkl} B_{bl} \tensor{E}{_{c} ^{i}}  \\ 
& +{\tfrac{1}{h}} c \tensor{h}{_{a} ^{\lambda}} (2\tensor{h}{^{b} _{l}} g_{k\lambda} \tensor{h}{^{c} _{j}} + \tensor{h}{^{c} _{l}} g_{k\lambda } \tensor{h}{^{b} _{j}}   \\ 
& - 2 \tensor{h}{^{b} _{\lambda}} {g_{kl}} \tensor{h}{^{c} _{j}} - \tensor{h}{^{c} _{\lambda}} {g_{kl}} \tensor{h}{^{b} _{j}} ) \varepsilon ^{jkm} \varepsilon ^{lin} B_{bm} B_{cn}.
\end{split}
\end{equation*} 
	
Note that the complication of the right hand side of the above two equations corresponds to the nonlinearity of the gravitational field equation. However, for a weak limit we have $\tensor{J}{_{a}^{\rho}} \approx 0$ and therefore,
\begin{equation*}
\begin{split}
& \nabla \cdot \vec{E}_{a} \approx 0, \\ & \frac{1}{c^{2}} \frac{\partial }{\partial t} \vec{E}_{a} +\nabla \times \vec{B}_{a} \approx 0. 
\end{split}
\end{equation*}

So far we have two equations that built the Maxwell-like equations. The other two equations should correspond to the divergence of gravitomagnetic fields, $\nabla \cdot \vec{B}_{a} ,$ and the curl of gravitoelectric fields (plus time derivative of gravitomagnetic fields), $\nabla \times \vec{E}_{a} +c\partial _{0} \vec{B}_{a} .$ They are equivalent to $\partial ^{\alpha } (h \tensor{S}{_{a} ^{\beta \gamma}} )+\partial ^{\beta} (h \tensor{S}{_{a} ^{\gamma \alpha}} )+\partial ^{\gamma } (h \tensor{S}{_{a} ^{\alpha \beta}} )$. In EM such quantity vanishes, it is the Bianchi identity, when we identify $h \tensor{S}{_{a} ^{\alpha \beta }} $ as the electromagnetic field strength. In teleparallel gravity, on the other hand, the field strength $h\tensor{S}{_{a} ^{\alpha \beta}}$ does not fulfill the Bianchi identity and hence 
\[\nabla \cdot \vec{B}_{a} \ne 0,\qquad \nabla \times \vec{E}_{a} +\frac{\partial \vec{B}_{a} }{\partial t} \ne 0. \] 
	
Since there are no equations for the derivative of field strength other then (\ref{eq:A}) we are not able to identify the right hand side of the above equations. This means that we could not define the Gauss' law for the gravitomagnetic field and the Faraday's law. Note that the quantity that defines the Bianchi identity is the torsion, $\tensor{T}{^{a} _{\alpha \beta}} $:
\begin{equation}
\partial _{\alpha } \tensor{T}{^{a} _{\beta \gamma}} +\partial _{\beta } \tensor{T}{^{a} _{\gamma \alpha}} +\partial _{\gamma } \tensor{T}{^{a} _{\alpha \beta}} =0.  \label{eq:C}
\end{equation}
However, since $\tensor{T}{^{a} _{\alpha \beta}} $ cannot be expressed in the field strength with the same indices \textit{$\alpha$} and \textit{$\beta$}, see equation (\ref{eq:B}), the left hand side of the above equation cannot be written into 
\begin{equation*}
\partial ^{\alpha } (h \tensor{S}{_{a} ^{\beta \gamma}} )+\partial ^{\beta} (h \tensor{S}{_{a} ^{\gamma \alpha}} )+\partial ^{\gamma } (h \tensor{S}{_{a} ^{\alpha \beta}} ).
\end{equation*}
\noindent Thus expressions like the Gauss' law for magnetic field and the Faraday's law are unable to be derived when the gravitoelectric and gravitomagnetic fields are defined according to (\ref{eq:A2}). Such a drawback leads us to propose a different definition of GEM as compared to [\citen{Spaniol2010}].

\section{A New Alternative of GEM in TEGR}
In the previous section, the Gauss' law and the Faraday's law can't be derived from the Bianchi identity of $h\tensor{S}{_{a}^{\rho \sigma}}$. So, a new proposal is needed to obtain the full set of Maxwell-like equations. In EM, the Maxwell's equation can be obtain with the help of field strength tensor and its dual tensor. The similarity between the Maxwell field strength $F_{\mu \nu}$ in terms of electromagnetic potentials and the Weitzenb\"{o}ck torsion in terms of tetrad fields (equation \eqref{eq:TT}) gives an alternative definition of GEM in teleparallel gravity. 

The new alternative is instead of using $ h \tensor{S}{_{a}^{\mu \nu}}$ to obtain Gauss-like law and Faraday-like law, we define the torsion $\tensor{T}{^{a}_{\mu \nu}}$, as the gravitoelectromagnetic field strength where the gravitoelectric field $E$ and gravitomagnetic field $B$ correspond, as usual, to $\tensor{T}{^{a}_{0\nu}}$ and $\tensor{T}{^{a}_{ij}}$, respectively:
\begin{equation}
\tensor{T}{^{a} _{\mu \nu}} =\left(\begin{array}{cccc} {0} & {\frac{\tensor{E}{^{a} _{x}}}{c} } & {\frac{\tensor{E}{^{a} _{y}}}{c} } & {\frac{\tensor{E}{^{a} _{z}}}{c} } \\ {-\frac{\tensor{E}{^{a} _{x}} }{c} } & {0} & {- \tensor{B}{^{a} _{z}}} & {\tensor{B}{^{a} _{y}}} \\ {-\frac{\tensor{E}{^{a} _{y}}}{c} } & {\tensor{B}{^{a} _{z}}} & {0} & {- \tensor{B}{^{a} _{x}}} \\ {-\frac{\tensor{E}{^{a} _{z}}}{c} } & {- \tensor{B}{^{a} _{y}}} & {\tensor{B}{^{a} _{x}}} & {0} \end{array}\right).   \label{eq:CC}
\end{equation}
\noindent  With this definition, the Bianchi identity (\ref{eq:C}) leads to the fulfillment of the Gauss-like law and the Faraday-like law:
\begin{equation}
\partial _{i} B^{ai} =0,\qquad \tfrac{1}{c} \epsilon^{ijk} \partial_{j} \tensor{E}{^{a}_{k}} + \partial_{0} B^{ai} = 0.      \label{eq:CD}
\end{equation}
	
The other two equations of the Maxwell's like equations are deducible from equation (\ref{eq:A}). To show this, one should write $\tensor{ S}{_{a}^{\mu \nu}}$, recalling (\ref{eq:D}) and (\ref{eq:E}), in terms of torsions:
\begin{equation} 
\begin{split}
\tensor{S}{_{a} ^{\mu \nu}} & =  {\tfrac{1}{2}} (\tensor{K}{^{\mu \nu} _{a}} - \tensor{h}{_{a} ^{\nu}} \tensor{T}{^{\alpha \mu } _{\alpha}} + \tensor{h}{_{a} ^{\mu}} \tensor{T}{^{\alpha \nu } _{\alpha}} \\ 
& =  {\tfrac{1}{2}}({\tfrac{1}{2}} (\tensor{T}{_{a} ^{\mu \nu}} + \tensor{T}{^{\nu \mu } _{a}} - \tensor{T}{^{\mu \nu } _{a}} ) 
- \tensor{h}{_{a} ^{\nu }} \tensor{T}{^{\alpha \mu } _{\alpha }} + \tensor{h}{_{a} ^{\mu }} \tensor{T}{^{\alpha \nu } _{\alpha }} ). 
\end{split}
\end{equation}

\noindent Thus we have
\begin{equation}
\begin{split}
\partial _{\mu } (h \tensor{S}{_{a} ^{\mu \nu}} ) & = {\tfrac{1}{2}} (\partial _{\mu } h) ({\tfrac{1}{2}} (\tensor{T}{_{a} ^{\mu \nu }} + \tensor{T}{^{\nu \mu } _{a}} - \tensor{T}{^{\mu \nu } _{a}}) 
- \tensor{h}{_{a} ^{\nu}} \tensor{T}{^{\alpha \mu} _{\alpha }} + \tensor{h}{_{a} ^{\mu }} \tensor{T}{^{\alpha \nu } _{\alpha }})  \\ 
& ={\tfrac{1}{2}} h ({\tfrac{1}{2}} (\partial _{\mu } \tensor{T}{_{a} ^{\mu \nu }} +\partial _{\mu } \tensor{T}{^{\nu \mu } _{a}} - \partial _{\mu } \tensor{T}{^{\mu \nu } _{a}}) \\
& \quad - \partial _{\mu } (\tensor{h}{_{a} ^{\nu}} \tensor{T}{^{\alpha \mu } _{\alpha }} )+ \partial _{\mu } (\tensor{h}{_{a} ^{\mu}} \tensor{T}{^{\alpha \nu } _{\alpha}}). 
\end{split} 
\end{equation}
The field equations become
\begin{equation}
\begin{split}
-{\frac{h}{2}} ({\partial _{\mu }} \tensor{T}{_{a} ^{\nu \mu}} & +\partial _{\mu } \tensor{T}{^{\mu \nu } _{a}} - \partial _{\mu } \tensor{T}{^{\nu \mu } _{a}}
- 2 \tensor{T}{^{\sigma \nu } _{\sigma}} \partial _{\mu } \tensor{h}{_{a} ^{\mu }}\\
& -2 \tensor{h}{_{a} ^{\mu }} \partial _{\mu } \tensor{T}{^{\sigma \nu } _{\sigma }} + 2 \tensor{T}{^{\sigma \mu } _{\sigma }} \partial _{\mu } \tensor{h}{_{a} ^{\nu }} 
+2 \tensor{h}{_{a} ^{\nu }} \partial _{\mu } \tensor{T}{^{\sigma \mu } _{\sigma }})  = kh \tensor{J}{_{a} ^{\nu}} \label{eq:G1}
\end{split}
\end{equation} 
with $k=8\pi G/c^{4} $. In order to derive Gauss-like and Ampere-like equations, we have to manipulate the indices in (\ref{eq:G1}). So, we have
\begin{equation}
\begin{split}
&-\frac{h}{2}( \partial _{\mu } (g^{\alpha \nu } g^{\beta \mu } \eta _{ab} \tensor{T}{^{b} _{\alpha \beta }} )+\partial _{\mu } (g^{\alpha \nu } \tensor{h}{_{b}^{\mu}} \tensor{h}{_{a}^{\beta}} \tensor{T}{^{b} _{\alpha \beta}} ) 
- \partial _{\mu } (g^{\alpha \mu } \tensor{h}{_{b}^{\nu}} \tensor{h}{_{a}^{\beta}} \tensor{T}{^{b} _{\alpha \beta}} )  \\
& -2 \tensor{h}{_{b}^{\sigma }} g^{\alpha \nu } \tensor{T}{^{b} _{\alpha \sigma}} \partial _{\mu } \tensor{h}{_{a} ^{\mu}}
- 2 \tensor{h}{_{a} ^{\mu}} \partial _{\mu } (\tensor{h}{_{b}^{\sigma}} g^{\alpha \nu } \tensor{T}{^{b} _{\alpha \sigma}}) 
+2\tensor{h}{_{b}^{\sigma}} g^{\alpha \mu } \tensor{T}{^{b} _{\alpha \sigma}} \partial _{\mu } \tensor{h}{_{a} ^{\nu}} \\
& \qquad \qquad \qquad +2 \tensor{h}{_{a} ^{\nu }} \partial _{\mu } (\tensor{h}{_{b}^{\sigma }} g^{\alpha \mu } \tensor{T}{^{b} _{\alpha \sigma }} )) = kh \tensor{J}{_{a} ^{\nu }}. \label{eq:AA}
\end{split}
\end{equation}

The lefthand side of equation (\ref{eq:AA}) consists of seven different terms. By  expanding every contraction on each term and using equation (\ref{eq:CC}) we get a long equation containing high powers of $h$-field. Considering a weak field approximation i.e. by disregarding terms of the form higher than quadratic power in tetrads we have, for $\nu =0$,  the Gauss-like equation for TEGR
	
\begin{equation}
\begin{split}
(g^{00} \eta _{ab} - 2 \tensor{h} {_{b} ^{0}}  \tensor{h} {_{a} ^{0}}) \partial ^{i} \tensor{E}{^{b}_{i}} \approx & -2k c \tensor{J}{_{a} ^{0}} 
+ ( g^{i0} \eta _{ab} - 2 \tensor{h} {_{b} ^{i}}  \tensor{h} {_{a} ^{0}}) \partial _{0} \tensor{E}{^{b}_{i}} \\ 
& + c ( g^{i0} \eta _{ab} - 2 \tensor{h} {_{b} ^{i}}  \tensor{h} {_{a} ^{0}}) \epsilon _{ijk} \partial ^{j} B^{bk} . \label{eq:DD}
\end{split}
\end{equation}
and, for $\nu=1,2,3$, the Ampere-like laws for TEGR
	
\begin{equation}
\begin{split}
( g^{in} \eta _{ab} -  2 \tensor{h} {_{b} ^{i}}  \tensor{h} {_{a} ^{n}}) (\partial ^{0} \tensor{E}{^{b}_{i}} + & c  \epsilon _{ijk} \partial ^{j} B^{bk}) \approx \\
& ( g^{0n} \eta _{ab} - 2 \tensor{h} {_{b} ^{0}}  \tensor{h} {_{a} ^{n}}) \partial ^{i} \tensor{E} {^{b}_{i}} + 2k c \tensor{J}{_{a} ^{n}} .  \label{eq:DF}
\end{split}
\end{equation}
	
\noindent So, now we have a full set of Maxwell-like equations under TEGR scheme, namely equations (\ref{eq:CD}), (\ref{eq:DD}) and (\ref{eq:DF}). In the next section, we will apply these Maxwell-like equations to obtain the gravitoelectromagnetic fields that corresponds to Schwarzschild metric and gravitational waves.
	
In the last two equations above, we haven't looked into detail the source terms. By recalling equations (\ref{eq:D}) and (\ref{eq:E}) we obtain that
\begin{equation}
\begin{split}
\tensor {J}{_{a}^{\rho}} = & \tfrac{1}{4 \pi G} ( \tfrac{1}{4} \tensor {h}{_{a}^{\lambda}} \tensor {T}{^{c}_{\mu \lambda}} (g^{\rho \gamma} g^{\alpha \mu} \eta _{cb} \tensor {T}{^{b}_{\alpha \gamma}}
+ ( g^{\rho \gamma} \tensor {h}{_{c}^{\mu}} - g^{\mu \gamma} \tensor {h}{_{c}^{\rho}}) \tensor {h}{_{b}^{\alpha}} \tensor {T}{^{b}_{\gamma \alpha}}) \\
& - \tfrac{1}{16} \tensor {h}{_{a}^{\rho}} \tensor {T}{^{c}_{\mu \nu}} (g^{\nu \gamma} g^{\alpha \mu} \eta _{cb} \tensor {T}{^{c}_{\alpha \gamma}}
+ ( g^{\nu \gamma} \tensor {h}{_{c}^{\mu}} - g^{\mu \gamma} \tensor {h}{_{c}^{\nu}}) \tensor {h}{_{b}^{\alpha}} \tensor {T}{^{b}_{\gamma \alpha}})). \label{eq:J}
\end{split}
\end{equation}
\noindent Every $\tensor {T}{^{a}_{\mu \nu}}$ contains one tetrad and every $g^{\mu \nu}$ contains two tetrads. Thus, we can see that every term in (\ref{eq:J}) contain seven tetrads. If we apply the weak field limit up to $(\tensor{h}{_{a}^{\mu}})^2$ we get
\begin{equation}
\tensor {J}{_{a}^{\rho}} \approx 0 \label{eq:J1}.
\end{equation}

\section{Examples}
\subsection{TEGR-GEM in Schwarzschild Spacetime}
Let us now look at TEGR-GEM in a Schwarzschild spacetime characterized by the following metric
\begin{equation}
\noindent {g_{\mu \nu }} = \begin{pmatrix}
			(1-\tfrac{2mG}{r})&0&0&0\\
			0&-(1-\tfrac{2mG}{r})^{-1}&0&0\\
			0&0&-r^{2}&0\\
			0&0&0&-r^{2} \sin ^{2} \theta
\end{pmatrix}.
\label{eq:FF}
\end{equation}
	
\noindent The above metric is defined in a spherical coordinate. Because the torsion (\ref{eq:CC}) is in Cartesian coordinate, we need to express the Schwarzschild metric in Cartesian coordinates (alternatively we may keep the metric in spherical coordinates but then  rewriting  the Maxwell-like equations in spherical coordinates are more tedious). One has
\begin{equation}
\noindent g_{\mu \nu} = \begin{pmatrix}
			(1-\tfrac{2mG}{r}) & 0 & 0 & 0\\
			0 & x^2 Q -1 & xyQ & xzQ\\
			0 & xyQ & y^2 Q-1 & yzQ\\
			0 & xzQ & yzQ & z^2 Q -1
\end{pmatrix},
\label{eq:FF1}
\end{equation}
where
\begin{equation}
Q(x,y,z) = \frac{-1}{\left( 1-\tfrac{2mG}{r}\right) r^2}
-\left( 1- \tfrac{z^2}{r^2}\right) \tfrac{z^2}{r^4} + 
\tfrac{1}{x^2+y^2}
\end{equation}
	
\noindent and $r^2=x^2+y^2+z^2$. The inverse of $g^{\mu \nu}$ are 
\begin{equation}
g^{\mu \nu} = \begin{pmatrix}
A^{-1} & 0 & 0 & 0\\
0 & \frac{1-(y^2+z^2)Q}{k} & \frac{xyQ}{k} & \frac{xzQ}{k}\\
0 & \frac{xyQ}{k} & \frac{1-(x^2+z^2)Q}{k} & \frac{yzQ}{k}\\
0 & \frac{xzQ}{k} & \frac{yzQ}{k} & \frac{1-(x^2+y^2)Q}{k}	\end{pmatrix},
\end{equation}

\noindent where
\begin{equation}
\begin{split}
k = & 3 x^2 y^2 z ^2 Q^3 - (x^2 y^2 + x^2 z^2 + y^2 z^2 
+x^2 yz + x y^2 z+xy z^2)Q^2 \\
& +(x^2 + y^2 + z^2 +xy + xz + yz)Q- 1
\end{split}
\end{equation}
and $A = 1 - \tfrac{2mG}{r} $.
	
The corresponding tetrad can be found by using the relation (\ref{eq:S1}). So, the tetrad for metric (	\ref{eq:FF1}) is
\begin{equation}
\tensor{h}{^{a} _{\mu }} = \begin{pmatrix}
\sqrt{AV} & x \sqrt{Q} & y \sqrt{Q} & z \sqrt{Q}\\
y \sqrt{AQV} & 0 & 1 & 0\\
z \sqrt{AQV} & 0 & 0 & 1\\
x \sqrt{AQV} & 1 & 0 & 0
\end{pmatrix}.
\label{eq:G}
\end{equation}
	
\noindent and its inverse $\tensor{h}{_{a}^{\mu}}$ is, 
\begin{equation}
{ \tensor{h}{_{a}^{\mu}}} = \begin{pmatrix}
\sqrt{\frac{V}{A}} & -y  \sqrt{\frac{QV}{A}} & -z \sqrt{\frac{QV}{A}} & -x \sqrt{\frac{QV}{A}} \\
- x V \sqrt{Q} & xyQW & xzQW & 1 + x^2 Q W\\
- y V \sqrt{Q} & 1+y^2 Q W & yzQW &  xyQW\\
- z V \sqrt{Q} & yzQW & 1+ z^2 QW & xzQW.
\end{pmatrix},
\end{equation}

\noindent with 
\begin{align}
& V(x,y,z) = \tfrac{1}{1 - r^2 Q},\\
& W(x,y,z) = \tfrac{1}{1 - r^2 \sqrt{Q}}.
\end{align} 
	
To obtain $\vec{E}^{a}$ and $\vec{B}^{a}$ fields we use the definitions of (\ref{eq:TT}) and (\ref{eq:CC}). Combining equations (\ref{eq:TT}), (\ref{eq:FF1}), and (\ref{eq:G}) gives components of torsion tensor $\tensor{T}{^{a} _{\mu \nu}}$. Considering all possible indices we obtain the non-zero gravitoelectric and gravitomagnetic fields
\begin{align}
& \vec{E}^0 = -c \nabla \sqrt{\tfrac{A}{1-r^2 Q}}, \label{eq:EG1}\\
& \vec{E}^1 = -c \nabla \left( y \sqrt{\tfrac{A Q}{1-r^2 Q}}\right) , \label{eq:EG2}\\
& \vec{E}^2 = -c \nabla \left( z \sqrt{\tfrac{A Q}{1-r^2 Q}}\right) , \label{eq:EG3}\\
& \vec{E}^3 = -c \nabla \left( x \sqrt{\tfrac{A Q}{1-r^2 Q}}\right) , \label{eq:EG4}\\
& \vec{B}^0 = \vec{r} \times \nabla \sqrt{Q}. \label{eq:BG1} 
\end{align}

The corresponding scalar potential $\Phi ^a$ and vector potential $\vec{A^a}$ are the following
\begin{equation}
\begin{split}
& \Phi^0 = c \sqrt{\frac{A}{1-r^2 Q}}, 
\quad \Phi^1 = c y \sqrt{\frac{AQ}{1-r^2 Q}}, \\
& \Phi^2 = c z \sqrt{\frac{AQ}{1-r^2 Q}}, 
\quad \Phi^3 = c x \sqrt{\frac{AQ}{1-r^2 Q}}, \\
& \vec{A}^0 = -\vec{r} \sqrt{Q}.
\end{split}
\end{equation}
As $\vec{B}^1=\vec{B}^2=\vec{B}^3=0$, we may choose vector potentials $\vec{A}^1=\vec{A}^2=\vec{A}^3=0$. So, we have a set of scalar potentials
\begin{equation}
(\Phi^0,\Phi^1,\Phi^2,\Phi^3) = \sqrt{\tfrac{A}{1-r^2 Q}}(1,y\sqrt{Q},z\sqrt{Q},x\sqrt{Q})
\end{equation}
and a set of vector potentials
\begin{equation}
(\vec{A}^0,\vec{A}^1,\vec{A}^2,\vec{A}^3) = (-\vec{r} \sqrt{Q},0,0,0).
\end{equation}

Finally, we note that all the fields are static, which are in agreement with the static nature of the Schwarzschild metric. Note also that we have both gravitoelectric and gravitomagnetic fields for $a=0$. They are not perpendicular one to the other. For $a=1,2,3$, however, we only have gravitoelectric fields.  
	
\subsection{TEGR-GEM in Gravitational Waves}	According to standard definition of gravitational waves [\citen{Weber2004,Misner1973}], vacuum gravitational fields equations satisfy	\begin{equation}
\square \overline{h}_{\mu \nu } = 0,    \label{eq:PA}
\end{equation}
	
\noindent where $\overline{h}_{\mu \nu } $ is the gravitational field defined in the linearized field theory.
	
The standard definition for gravitational waves are designed for transverse-traceless (TT) gauge [\citen{Misner1973}] which means that
\begin{equation}
	h=h_{\alpha }^{\; \alpha }= 0.
\end{equation}
The TT-gauge requires the non-zero components for  $h_{\mu \nu } $ as follows
\begin{equation}
\begin{split}
h_{11} =-h_{22} =A(t,z) = Re(A_{+} e^ {-i \omega (t-z)} )\\ 
h_{12} = h_{21} =C(t,z) =  Re(A_{\times} e^ {-i \omega (t-z)} ).  \label{eq:V}
\end{split}
\end{equation}
\noindent In the above $A_{+} (z)$ and $A_{\times} (z)$ are the amplitudes of the solutions of equation (\ref{eq:PA}) which correspond to the two  independent modes of polarization ($"+"$ and $"\times"$)  for gravitational waves propagating along the z-axis. 
	
The metric tensor for gravitational wave according to equations (\ref{eq:PA}) and (\ref{eq:V}) is
\begin{equation}
\noindent {g_{\mu \nu }} = \begin{pmatrix}
1&0&0&0\\
0&{A-1}&C&0\\
0&C&{-A-1}&0\\
0&0&0&-1
\end{pmatrix}
\label{eq:V1}
\end{equation}
\noindent and its inverse is
	
\begin{equation}
\noindent {g^{\mu \nu }} = \begin{pmatrix}
1&0&0&0\\ 
0&\frac{-A-1}{1-A^{2} -C^{2} }&\frac{-C}{1-A^{2} -C^{2}}&0 \\
0&\frac{-C}{1-A^{2} -C^{2} } & \frac{A-1}{1-A^{2} -C^{2} }&0\\ 
0&0&0&-1. 
\end{pmatrix}.
\label{eq:V2}
\end{equation}
	
\noindent Here we used the Minkowski tensor with signature $(1,-1,-1, -1)$. Using (\ref{eq:S1}) and (\ref{eq:S2}) we obtain the tetrad 
\begin{equation}
{ \tensor{h}{^{a}_{\mu}}} = \begin{pmatrix}
1 & 0 & 0 & 0\\
0 & \frac{1}{2} \left( \Delta - \frac{2A}{\Delta}\right) & - \frac{C}{\Delta} & 0 \\
0 &- \frac{C}{\Delta} & \frac{1}{2} \left( \Delta + \frac{2A}{\Delta}\right) & 0 \\
0 & 0 & 0 & -1
\end{pmatrix},
\label{eq:Z4}
\end{equation}
and its inverse 
	
\begin{equation}
{ \tensor{h}{_{a}^{\mu}}} = \begin{pmatrix}
1 & 0 & 0 & 0\\
0 &  \tfrac{1}{2 \delta} \left(\Delta + \tfrac{2A}{\Delta}\right) & \frac{C }{\Delta \delta}  & 0 \\
0 & \frac{C}{\Delta \delta}  &  \tfrac{1}{2 \delta} \left( \Delta - \tfrac{2A}{\Delta}\right) & 0 \\
0 & 0 & 0 & -1
\end{pmatrix},
\end{equation}
	
\noindent where
\begin{align}
& \Delta (A,C) = \sqrt{2+2 \sqrt{1 - A^2 - C^2}},\\
& \delta = \tfrac{1}{4} \left(\Delta^2 - \tfrac{4A^2}{\Delta^2}\right) - \left( \tfrac{C^2}{\Delta^2}\right).
\end{align}
	
Using (\ref{eq:Z4}) and recalling (\ref{eq:S1}) and (\ref{eq:CC}) we can obtain $E^{a} $ and $B^{a} $ fields. The non-zero components of $E^{a} $ and $B^{a} $ are
\begin{align}
& \tensor{E}{^{1} _{x}} =\frac{c}{2} \partial_{0} \left( \Delta - \frac{2A}{\Delta}\right) \label{eq:EG13},\\
& \tensor{E}{^{2} _{x}} =-c \partial_{0} \left( \frac{C}{\Delta}\right) \label{eq:EG14} ,\\
& \tensor{E}{^{1} _{y}} =-c \partial_{0} \left( \frac{C}{\Delta}\right) \label{eq:EG15},\\
& \tensor{E}{^{2} _{y}} = \frac{c}{2} \partial_{0} \left( \Delta + \frac{2A}{\Delta}\right) \label{eq:EG16},\\
& \tensor{B}{^{1} _{x}} =-\partial_{3} \left( \frac{C}{\Delta}\right) \label{eq:BG4}, \\
& \tensor{B}{^{2} _{x}} = \frac{1}{2} \partial_{3} \left( \Delta + \frac{2A}{\Delta}\right) \label{eq:BG5},\\
& \tensor{B}{^{1} _{y}} = - \frac{1}{2} \partial_{3} \left( \Delta - \frac{2A}{\Delta}\right) \label{eq:BG6}, \\
& \tensor{B}{^{2} _{y}} = \partial_{3} \left( \frac{C}{\Delta}\right) \label{eq:BG7}.
\end{align}
It turns out that the fields are transversal as $\tensor{E}{^{a}_{z}}=\tensor{B}{^{a}_{z}}=0$. However, $\vec{E^a}$ and $\vec{B^a}$ are not perpendicular one another except for the case when the amplitude $A_{+}$ and $A_{\times}$ are very small (weak field case) as in this case
\begin{align}
& \partial_{0} \Delta \rightarrow 0. 
\end{align}
	
We may derive the corresponding potentials. Considering the vector potentials $\vec{A}^a$ depend only on time $t$ and coordinate $z$, the expression $\vec{B}^a = \nabla \times \vec{A}^a$ and equations (\ref{eq:BG4} - \ref{eq:BG7}) give
\begin{align}
& \vec{A}^1 = \left (\frac{A}{\Delta}-\frac{\Delta}{2}, \frac{C}{\Delta}, 0 \right ) \label{eq:VP1}\\
& \vec{A}^2 = \left (\frac{C}{\Delta}, - \frac{A}{\Delta}-\frac{\Delta}{2}, 0 \right ) \label{eq:VP2}
\end{align}
	
On the other hand, recalling the definition of scalar potential $\Phi ^a$ through 
\begin{equation}
\vec{E}^{a} =-c\nabla \Phi ^{a} -\frac{\partial \vec{A}^{a} }{\partial t} 
\end{equation}
the gravitoelectric fields (\ref{eq:EG13}-\ref{eq:EG16}) and the vector potentials (\ref{eq:VP1} - \ref{eq:VP2}) gives $\partial_x \Phi^1 = \partial_y \Phi^1 =\partial_z \Phi^1 =0$, i.e. $\Phi^1$ is space coordinate independent, and similarly for $\Phi^2$.
	
In EM theory, Poynting vector represents the rate of energy transfer per unit area. It corresponds to terms containing both electric and magnetic fields in the canonical stress tensor derived from EM Lagrangian [\citen{Jackson1999,Barnett2014}]. Analogously we take into account the stress tensor for GEM
\begin{equation}
\tensor{\Theta}{^{\mu} _{\lambda}} = \frac{\partial \mathfrak{L}}{\partial(\partial _{\mu} \tensor{b}{^{a} _{\nu} })} \partial _ {\lambda} \tensor{b}{^{a} _{\nu}} - \tensor{\delta}{^{\mu} _{\lambda}} \mathfrak{L},
\end{equation}
with the Lagrangian is given by (\ref{eq:L}). By considering the weak field approximation, the first term of (\ref{eq:H}) may be replaced by the Kronecker delta. Accordingly the canonical stress tensor reads
\begin{equation}
\begin{split}
\tensor{\Theta}{^{\mu} _{\lambda}} = & \frac{1}{2k} (\tensor{T}{^{a} _{\lambda \nu}} - \partial_\lambda \tensor{b}{^{a} _{\nu}}  )  [   \tfrac{1}{2} \tensor{T}{_{a} ^{\mu \nu}} + \tfrac{1}{2} \tensor{\delta}{_{b} ^{\nu}} \tensor{\delta}{^{e} _{\beta}} \eta _{ea} \eta ^{bc} \tensor{T}{_{c} ^{\mu \beta}}  -  \tfrac{1}{2} \tensor{\delta}{_{b} ^{\mu}} \tensor{\delta}{^{e} _{\beta}} \eta _{ea} \eta ^{bc} \tensor{T}{_{c} ^{\nu \beta}} \\
& \qquad \qquad \qquad \qquad  - \tensor{\delta}{_{a} ^{\nu}} \tensor{\delta}{^{e} _{\beta}} \tensor{\delta}{_{e} ^{c}} \tensor{T}{_{c} ^{\mu \beta}} + \tensor{\delta}{_{a} ^{\mu}} \tensor{\delta}{^{e} _{\beta}} \tensor{\delta}{_{e} ^{c}} \tensor{T}{_{c} ^{\nu \beta}} ]  \\
&  - \tensor{\delta}{^{\mu} _{\lambda}} \frac{1}{2k} 
[ \tfrac{1}{4} \eta _{bc} \eta ^{gh}
\tensor{\delta}{_{g} ^{\beta}} \tensor{\delta}{_{h} ^{\nu}}  \tensor{T}{^{b}  _{0 \nu}} \tensor{T}{^{c} _{0 \beta}} 
+ \tfrac{1}{4} \eta _{bc} \eta ^{gh} \eta ^{ij}
\tensor{\delta}{_{g} ^{\beta}} \tensor{\delta}{_{h} ^{\nu}}  \tensor{T}{^{b} _{i \nu}} \tensor{T}{^{c} _{j \beta}} \\
& \qquad \qquad + \tfrac{1}{2} 
\tensor{\delta}{_{b} ^{\lambda}} \tensor{\delta}{_{c} ^{\nu}}  \tensor{T}{^{b} _{0 \nu}} \tensor{T}{^{c} _{0 \lambda}} 
+ \tfrac{1}{2} 
\tensor{\delta}{_{b} ^{\lambda}} \tensor{\delta}{_{c} ^{\nu}} \eta^{ij} \tensor{T}{^{b} _{i \nu}} \tensor{T}{^{c} _{j \lambda}} \\
& \qquad \qquad - \tensor{\delta}{_{b} ^{\lambda}} \tensor{\delta}{_{c} ^{\nu}}  \tensor{T}{^{b} _{0 \lambda}} \tensor{T}{^{c} _{0 \nu}}
- \tensor{\delta}{_{b} ^{\lambda}} \tensor{\delta}{_{c} ^{\nu}} \eta^{ij} \tensor{T}{^{b} _{i \lambda}} \tensor{T}{^{c} _{j \nu}}].
\end{split}
\end{equation}
By expanding all combination of contraction over torsion tensor, several candidate terms for GEM Poynting vector are obtained. Here we only consider those terms. Next, considering the non-zero fields of (\ref{eq:EG13} - \ref{eq:BG7}) we have, after multiplying with some constants to adjust the units,
\begin{equation}
\begin{split}
\tensor{S}{^3 _0} = & \frac{1}{4k} [ (\tensor{E}{^2 _x} \tensor{B}{^1 _x} - \tensor{E}{^1 _x} \tensor{B}{^2 _x})
+  (\tensor{E}{^2 _y} \tensor{B}{^1 _y} - \tensor{E}{^1 _y} \tensor{B}{^2 _y})] \\
& + \frac{1}{2k} [ (\tensor{E}{^1 _x} \tensor{B}{^1 _y} - \tensor{E}{^1 _y} \tensor{B}{^1 _x})
+ (\tensor{E}{^2 _x} \tensor{B}{^2 _y} - \tensor{E}{^2 _y} \tensor{B}{^2 _x})]. \label{eq:P1}
\end{split}
\end{equation}
Unlike EM theory, here the Poynting vector is found from the $\tensor{\Theta}{^3 _0}$ not from $\tensor{\Theta}{^0 _0}$ components. 

The Poynting vector in EM theory is the cross product of $E$ and $B$ fields in the same spacetime, but here we found an interesting forms. From (\ref{eq:P1}) we have two kinds of cross product, one in the flat spacetime index (the first and second terms) and other in curved spacetime index (the third and forth terms). These two kinds of cross products reveals the gravitational waves energy transfer both in the flat spacetime (tangent space) and in the curved spacetime. Finally, applying (\ref{eq:EG13} - \ref{eq:BG7}), we get
\begin{equation}
\begin{split}
\tensor{S}{^3 _0} = & \frac{c}{2k} \left(  - \partial_{0} \left( \frac{C}{\Delta}\right) 
\partial_{3} \left( \frac{C}{\Delta}\right) - 
\frac{3}{4} \partial_{0} \Delta
\partial_{3}\Delta 
- \frac{1}{4} \partial_{0} \left( \frac{2A}{\Delta}\right)
\partial_{3} \left( \frac{2A}{\Delta}\right) \right) . \label{eq:P3}
\end{split}
\end{equation}

\section{Conclusions}
Theory of GEM was introduced due to the similar expression between the Coulomb force and the Newton gravitational force. Accordingly, as the EM is governed by the Maxwell’s equations, the Newtonian GEM is governed by the Maxwell-like gravitational equations. In addition, we have two versions of the Maxwell-like equations in Einstein gravity for the weak limit case. Here, we have also showed the Maxwell-like equations in TEGR.
	
The first proposal of GEM in TEGR was given by Spaniol and Andrade [\citen{Spaniol2010}] where the superpotentials (times a factor of $h$) $h \tensor{S}{_{a}^{\rho \sigma}}$ are taken to play the role as the field strengths because $h \tensor{S}{_{a}^{\rho \sigma}}$ fulfill equations (\ref{eq:A}) which look like the inhomogeneous Maxwell equations. Accordingly, the inhomogeneous parts of the Maxwell-like equations are obtained. However, as $h \tensor{S}{_{a}^{\rho \sigma}}$ do not fulfill the Bianchi identity and thus the homogeneous parts of the Maxwell-like equations could not be derived. This drawback leads us to propose different definition of gravitoelectromagnetic fields in TEGR.  Because of similar expressions between the Weitzenb\"{o}ck torsion in terms of tetrad fields and the Maxwell field strength in terms of electromagnetic potentials we then propose the Weitzenb\"{o}ck torsion as the gravitational field strength. Just like in EM, gravitoelectric and gravitomagnetic fields are then defined from the field strength components. The gravitomagnetic Gauss' law and the Faraday-like law, i.e. the homogeneous parts of the Maxwell-like equations, are automatically obtained. The inhomogeneous parts of the Maxwell-like equations are obtained from equations (\ref{eq:A}) after writing the superpotential in terms of torsions, or equivalently in terms of gravitoelectric and gravitomagnetic fields, and after taking some weak field approximations.
	
We consider two examples of deriving  gravitoelectromagnetic fields, first is the gravitoelectric fields in the Schwarzschild spacetime and the second is that in gravitational waves. For the Schwarzschild spacetime, we obtain as expected that the gravitoelectromagnetic fields are static. For Minkowski spacetime index $a=0$, both the gravitoelectric and the gravitomagnetic fields exist and they are not perpendicular one to the other. There are no gravitomagnetic fields for $a=1,2,3$. For each index $a$ we obtain a pair of scalar and vector potentials but with zero vector potentials for $a=1,2,3$. For the second case we consider the gravitational waves propagating along $+z$ direction. We obtain that there are gravitoelectric as well as gravitomagnetic fields only for Minkowski spacetime indices $a=1,2$. They are transversal but are not perpendicular one to the other except for very small amplitudes. The corresponding scalar potentials and the $z$-components of vector potentials are zero. Finally we derived the Poynting vector from the stress tensor. It consists of cross-products both in Minkowski spacetime as well as in curved spacetime between gravitoelectric and gravitomagnetic fields.

\section*{Acknowledgments}
This research was financially supported by the \textit{Riset dan Inovasi KK Institut Teknologi Bandung}, and \textit{Desentralisasi DIKTI}. Kian Ming was supported by \textit{Beasiswa Unggulan} and I-MHERE FMIPA ITB.


\end{document}